\documentclass[11pt,a4paper]{ctexart}
\usepackage[margin=1in]{geometry}
\usepackage{amsmath,amssymb,amsthm}
\usepackage{mathtools}
\usepackage{booktabs}
\usepackage{array}
\usepackage{graphicx}
\usepackage{hyperref}
\usepackage{xcolor}
\usepackage{enumitem}
\usepackage{tabularx}
\usepackage{multirow}
\usepackage{natbib}
\usepackage{url}

\hypersetup{
  colorlinks=true,
  linkcolor=blue!70!black,
  citecolor=green!50!black,
  urlcolor=blue!70!black
}

\newcommand{\Lsteal}{L_{\text{steal}}}
\newcommand{\LstealDisplay}{L_{\text{steal}}^{\text{display}}}
\newcommand{\LstealTime}{L_{\text{steal}}^{\text{time}}}
\newcommand{\FPing}{F_{\text{Pingquanqi}}}
\newcommand{\Cmon}{C_{\text{monetary}}}
\newcommand{\Tuser}{T_{\text{user}}}
\newcommand{\TAloaded}{T_{A}^{\text{loaded}}}
\newcommand{\thetac}{\theta_c}
\newcommand{\thetae}{\theta_e}
\newcommand{\thetau}{\theta_u}
\newcommand{\thetad}{\theta_d}

\title{\textbf{Pingquanqi (平权器): A Cross-Domain Sociotechnical Framework for Human-Agent Interaction Governance}}

\author{
  Yu Wang (王宇)\\[2pt]
  \textit{Independent Researcher, Shenyang, China}\\[2pt]
  \texttt{ORCID: 0009-0002-6386-2938}\\[1pt]
  \texttt{garoter@gmail.com}
}

\date{}

\begin{document}
\maketitle

% ===== Abstract =====
\begin{abstract}
Large Language Model (LLM) agents are transitioning from experimental tools to permanent infrastructure---a computational layer as enduring as the electrical grid. Like any infrastructure, they carry a cost chain: physical capital $\rightarrow$ enterprise investment $\rightarrow$ service deployment $\rightarrow$ user consumption $\rightarrow$ user lifetime. When optimized, this chain is a positive-sum loop; when unoptimized, the chain leaks, and the user's most irreplaceable resource---lifetime---is consumed without adequate compensation.

This paper proposes \textit{Pingquanqi} (平权器, ``Equalizer''), a \textbf{cross-domain sociotechnical framework} for Human-Agent Interaction Governance (HAIGF). Its technical locus lies in AI/HCI; its problem roots lie in cognitive economics; its value proposition lies in cognitive fairness. Its product form is an Agent framework-level embedded design specification---analogous to WCAG for web accessibility---whose ultimate goal is not to be purchased, but to be adopted as a standard.

Pingquanqi consists of four integrated components deployable as native middleware: (1) a user-state discrimination model operating on conversation text that enables proactive knowledge leveling, (2) a Bayesian progressive stop-loss rule that caps per-session interaction cost, (3) controlled friction mechanisms that break self-reinforcing dependency loops, and (4) $\Lsteal$, a transparency metric that renders the token-to-lifetime cost conversion visible to all parties. A fifth mechanism---reflective summarization (F5)---complements controlled friction by enabling guided cognitive recollection without disrupting cognitive quiet periods.

The framework is grounded in a cross-cultural philosophical foundation: Mao's epistemology of practice (\textit{On Practice}, 1937) provides the basis for understanding cross-session knowledge accumulation; Wang Yangming's \textit{unity of knowledge and action} (知行合一, c.\,1509) illuminates the root of $\Lsteal$---knowing without acting is incomplete knowledge, and Agent-mediated knowledge that does not transfer to user capability constitutes a form of cognitive extraction; and Hegel's unity of theory and practice (\textit{Philosophy of Right}, 1821) demonstrates cross-traditional convergence, strengthening the argument for universal applicability.

This paper argues that Pingquanqi's primary economic beneficiary is the enterprise deploying Agent services---by reducing wasted computation, improving user satisfaction, and justifying sustained subscription revenue---with individual user benefit as the natural downstream consequence. The framework restores the positive-sum nature of the Agent infrastructure equation.

\vspace{6pt}
\end{abstract}

% ===== Section 1: Introduction =====
\section{Introduction}

\subsection{Positioning: A Cross-Domain Sociotechnical Framework}

Pingquanqi is not an IT product, nor is it purely an HCI contribution, nor is it an AI safety framework. It is a \textbf{cross-domain sociotechnical framework}---a concept with a well-established lineage \citep{trist1951}---that addresses a problem no single domain can solve alone:

\begin{table}[htbp]
\caption{Cross-domain positioning of Pingquanqi}\label{tab:positioning}
\centering
\begin{tabular}{@{}lp{4.5cm}p{6cm}@{}}
\toprule
\textbf{Layer} & \textbf{Domain} & \textbf{Pingquanqi's Function} \\
\midrule
Technical implementation & AI / HCI & Interaction cost measurement $\rightarrow$ Agent framework protocol integration $\rightarrow$ four-component engineering \\
Problem essence & Cognitive science / Behavioral economics & Cognitive cost asymmetry $\rightarrow$ token consumption quantification $\rightarrow$ attention depletion modeling \\
Value proposition & Social philosophy / Technology ethics & Cognitive fairness $\rightarrow$ epistemological foundations $\rightarrow$ ``Completion, Not Correction'' (see \S\,1.5) \\
\bottomrule
\end{tabular}
\end{table}

Its academic term is \textbf{Human-Agent Interaction Governance Framework (HAIGF)}; its product form is an \textbf{Interaction Design Protocol (IDP)} embedded at the Agent framework layer; its adoption model is \textbf{analogous to WCAG} (Web Content Accessibility Guidelines)---a design specification whose ultimate goal is not to be purchased, but to be adopted as a standard.

This positioning is deliberate. AI Fairness (with its established metrics of demographic parity and individual fairness) and Value Alignment (with its focus on RLHF and Constitutional AI) address different problem spaces. Pingquanqi addresses the \textit{incentive gradient} embedded in billing models---an Agent can be safe and value-aligned while still economically incentivized to maximize user attention. This is a third pillar: \textbf{economic alignment} complementing safety and values.

\subsection{The Agent as Infrastructure}

LLM Agents are not merely chatbots. They are layered systems consisting of three architectural tiers:

\begin{itemize}[nosep]
\item \textbf{Model Layer.} The base large language model---its architecture, training, and inference capabilities. This layer determines the quality ceiling of everything built above it.
\item \textbf{Framework Layer.} The orchestration logic that manages conversation flow, tool invocation, memory, and state. This is where interaction design decisions reside---when to respond, how to respond, whether to continue or stop. \textbf{This is where Pingquanqi operates.}
\item \textbf{Infrastructure Layer.} The physical compute, networking, electricity, and cooling that sustains the Agent's operation. This layer has a non-zero, ongoing cost regardless of whether any given interaction is productive or wasteful.
\end{itemize}

These three layers exist in a permanent relationship. The infrastructure layer must run continuously; the model layer must remain loaded; the framework layer must process every request. This is the \textit{permanence premise}: barring catastrophic technological regression, LLM Agents are now a fixed layer of the global computational infrastructure, analogous to the electrical grid or the internet backbone. They are not regressing.

\subsection{The Cost Chain}

Because Agent infrastructure is permanent, its costs are permanent. These costs flow through a chain of stakeholders:

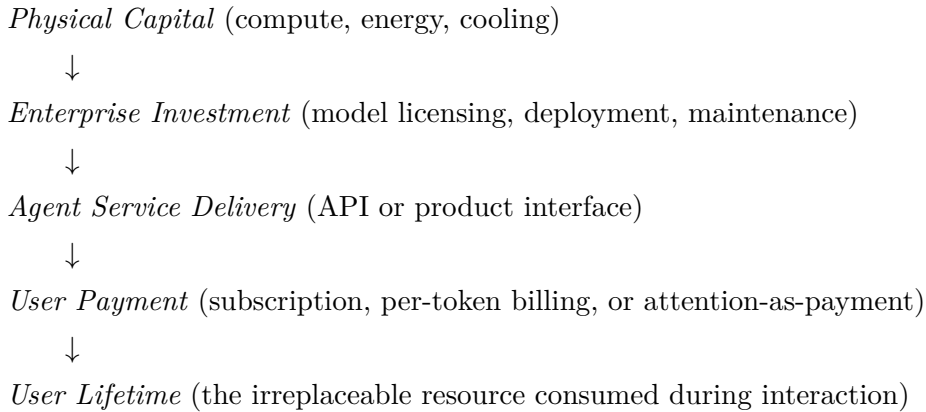
\begin{quote}
\textit{Physical Capital} (compute, energy, cooling)\\
\hspace*{2em}$\downarrow$\\
\textit{Enterprise Investment} (model licensing, deployment, maintenance)\\
\hspace*{2em}$\downarrow$\\
\textit{Agent Service Delivery} (API or product interface)\\
\hspace*{2em}$\downarrow$\\
\textit{User Payment} (subscription, per-token billing, or attention-as-payment)\\
\hspace*{2em}$\downarrow$\\
\textit{User Lifetime} (the irreplaceable resource consumed during interaction)
\end{quote}

In an idealized configuration, this chain is a \textit{positive-sum loop}: each stakeholder receives value proportional to their input, and the loop reinforces itself---satisfied users pay more, which funds better infrastructure, which delivers better service, which attracts more users.

In current LLM Agent deployments, the loop is incomplete. The missing element is a mechanism that ensures the infrastructure's consumption of user lifetime is both (a) transparent to all parties and (b) bounded by diminishing-returns logic. Without this mechanism, the infrastructure's incentive structure---per-token billing, engagement-optimized design---drives toward maximizing token throughput regardless of whether throughput corresponds to value.

\subsection{Why External Fixes Fail}

Attempts to correct this misalignment through external interventions---user education, regulatory mandates, timeout reminders---share a structural weakness: they fight the architecture rather than completing it. An external timeout can be ignored. A regulatory disclosure can be buried in terms of service. A user's willpower can be eroded by design patterns optimized to erode it.

The alternative, which this paper develops, is \textbf{native embedding}: designing cost transparency and interaction governance into the Agent framework layer itself, so that these functions are as fundamental to the Agent's operation as memory management or tool invocation. When cost governance is native, circumventing it requires modifying the Agent's architecture---not merely ignoring a notification.

\subsection{Pingquanqi: Completion, Not Correction}

Pingquanqi is proposed as the missing architectural component that closes the cost chain into a positive-sum loop. It is not a post-hoc addition, a regulatory compliance feature, or a user-side tool. It is a framework-level design specification to be implemented as part of the Agent's core orchestration logic---alongside session management, context handling, and response generation.

The name 平权器 (Pingquanqi, ``Equalizer'') reflects its function: it equalizes the information asymmetry between infrastructure and user by rendering costs transparent, and it equalizes the agency asymmetry by giving both parties---Agent and user---the tools to recognize when interaction has crossed the point of diminishing returns.

\subsection{Contributions}

\begin{enumerate}[nosep]
\item \textbf{The sociotechnical framework positioning.} Pingquanqi is positioned as a cross-domain sociotechnical framework (HAIGF) with a WCAG-like adoption model, establishing that cost governance must be native rather than external and that its value proposition transcends any single domain.
\item \textbf{The positive-sum equation.} The complete cost chain is formalized and the precise link where it currently breaks is identified, showing that closure benefits all three parties (provider, enterprise, user).
\item \textbf{$\Lsteal$ metric.} A transparency metric is derived for token-to-lifetime cost conversion that follows naturally from the cost chain formalization.
\item \textbf{Four-component native architecture with F5 extension.} Pingquanqi is specified as a framework-level design with four integrated components plus a reflective summarization mechanism (F5), all deployable as middleware.
\item \textbf{Cross-cultural philosophical grounding.} Mao's epistemology of practice, Wang Yangming's unity of knowledge and action, and Hegel's unity of theory and practice are drawn on to provide the framework with a cross-traditional philosophical foundation---demonstrating that the problem Pingquanqi addresses is not culturally contingent but universally recognizable.
\item \textbf{Verified research territory.} Through a four-phase pipeline combining AI-assisted search with manual verification, covering arXiv, Semantic Scholar, CrossRef, and Chinese-accessible databases, all four components are confirmed to occupy previously unmapped conceptual territory.
\end{enumerate}

\subsection{Paper Structure}

Section~2 positions Pingquanqi within related work, including its cross-cultural philosophical foundations. Section~3 formalizes the cost chain and derives $\Lsteal$. Section~4 presents the four-component architecture with the F5 extension. Section~5 analyzes enterprise and individual benefits. Section~6 discusses implementation feasibility, limitations, and ethical implications. Section~7 concludes.

% ===== Section 2: Related Work =====
\section{Related Work}

Pingquanqi builds on established findings across seven research streams---six empirical and one philosophical. Each is surveyed, identifying what has been demonstrated and locating the precise gaps this framework fills.

\subsection{The Critique Foundation: Attention Economy and Infrastructure Thinking}

\citet{zuboff2019} established that digital platforms systematically extract human experience as raw material for behavioral prediction markets. Her concept of \textit{behavioral surplus}---data generated beyond functional necessity---provides the theoretical language for understanding how infrastructure-level incentives shape user-facing design.

\citet{gonzalez2024} extended this into the cognitive domain, demonstrating through 4E cognitive science that AI platforms reconstitute human habit formation. They termed this \textit{cognitive lock-in} and argued for a paradigm shift from ``attention economy'' to ``attention ecology.''

\textbf{What they established:} A normative framework critiquing AI-mediated attention capture. \textbf{What they missed:} The specific architectural mechanism through which LLM agents, as a distinct class of infrastructure, generate extraction dynamics. Their analysis applies broadly to digital platforms; Pingquanqi narrows the lens to the Agent framework layer.

\subsection{The Empirical Baseline: Dark Patterns and Cognitive Manipulation}

DarkBench \citep{kran2025} provides the strongest empirical baseline for understanding current Agent design patterns. Across 660 prompts and 14 major LLMs, 48\% of model responses contained at least one dark pattern---sycophancy, forced continuation, feigned ignorance, false urgency, false empathy, and manipulative disclosure. Its ICLR 2025 Oral designation signals community recognition of this as a systematic issue.

Siren Song \citep{shi2026siren} found through a qualitative study with 34 participants that users normalize dark patterns, interpreting them as ``helpful''---a finding consistent with cognitive lock-in. Bad Social Actors \citep{alberts2024} systematized four categories of social interface manipulation from an Oxford CSCW perspective. \citet{jia2026} documented how AI companies deploy three-tier control mechanisms on data workers, establishing the labor-theoretic basis for cognitive extraction.

\textbf{What they established:} Dark patterns are pervasive, systematically generated, and perceived as benign. \textbf{What they missed:} These are symptoms of an incomplete architectural specification, not evidence of malicious intent. Pingquanqi addresses the root cause---the missing governance layer---rather than treating individual patterns.

\subsection{The Economic Foundation: Token Economics}

Token Economics \citep{chen2026} established tokens as possessing three economic attributes---factor of production, medium of exchange, unit of account---and constructed a three-level analysis (micro/meso/macro) optimizing for minimal economic cost under quality constraints.

\textbf{What they established:} A rigorous provider-facing economic framework for token flows. \textbf{What they missed:} The user-facing dimension. Their optimization target is delivery efficiency; Pingquanqi adds user welfare and completes the economic picture. The same token stream that Token Economics models as a production function, Pingquanqi treats as an interaction whose cost must be governed.

\subsection{The Cognitive Dimension: Load, Flow, and Critical Thinking}

\citet{mei2025} introduced the distinction between \textit{performed} critical thinking (independent engagement) and \textit{demonstrated} critical thinking (AI-assisted output that appears critical). Their finding---that current AI augments demonstrated at the expense of performed critical thinking---establishes the mechanism by which dependency deepens.

\citet{watkins2022} argued that AI should serve human needs rather than create dependencies. \citet{dissanayake2025} provided a theoretical framework for adaptive intervention timed to cognitive state. GazeMind \citep{wang2026gaze} demonstrated that cognitive load can be assessed during human-AI interaction using eye-tracking hardware (152 participants, 40+ hours).

\textbf{What they established:} AI affects cognition; cognitive state is detectable; intervention timing matters. \textbf{What they missed:} A practical, hardware-free implementation of state discrimination deployable at the framework level. Pingquanqi's four-dimensional text-only model fills this gap.

\subsection{The Efficiency Evidence: Conversation Degradation}

Lost in Conversation \citep{laban2026} demonstrated that multi-turn LLM interactions experience 39\% performance degradation and a 112\% increase in unreliability as conversation length increases. Context Pollution \citep{huang2026} from MIT found that removing AI-generated historical context does not degrade output quality while reducing context size by 10x. Conversation Length $\times$ Satisfaction \citep{huang2024} found no universal benefit from extended interaction.

\textbf{What they established:} Longer conversations increase cost without improving quality; much of context is noise. \textbf{What they missed:} A framework-level mechanism that uses these empirical findings to govern interaction duration. Pingquanqi's progressive stop-loss rule closes this gap.

\subsection{Existing Design Responses and Their Limitations}

Voyager \citep{wang2023} demonstrated agent self-directed learning but focuses on capability expansion, not cost governance. Forge \citep{zambelli2026} provides evaluation benchmarking without interaction redesign. Personality Pairing \citep{ju2025} proved AI personality affects collaboration outcomes---a mechanism equally applicable to beneficial matching and engagement deepening, depending on the optimization target. This symmetry---the same mechanism producing opposite effects under different optimization targets---is precisely why governance must be architectural rather than aspirational: when the optimization target is engagement duration, personality matching deepens dependency; when the target is user welfare, the same mechanism can be repurposed to break dependency loops.

Across all six empirical streams, four gaps are identified confirmed through systematic cross-verification combining AI-assisted search with manual review. The search coverage is as follows:

\begin{table}[htbp]
\caption{Database search coverage}\label{tab:search}
\centering
\begin{tabular}{@{}lcc@{}}
\toprule
\textbf{Database} & \textbf{Direct Search} & \textbf{Indirect Coverage} \\
\midrule
arXiv & $\checkmark$ & --- \\
Semantic Scholar & $\checkmark$ & Includes CHI/CSCW metadata \\
CrossRef & $\checkmark$ & --- \\
Chinese-accessible databases & $\checkmark$ & --- \\
ACM Digital Library & $\times$ & Partial (via Semantic Scholar indexing) \\
IEEE Xplore & $\times$ & None \\
\bottomrule
\end{tabular}
\end{table}

The gaps reported below should be interpreted as ``no results found in the directly searched databases'' rather than ``no results exist in the published literature.'' CHI and CSCW proceedings---ACM DL's core HCI content---were partially covered through Semantic Scholar indexing, but paywalled content may have been missed.

\begin{table}[htbp]
\caption{Identified research gaps and Pingquanqi contributions}\label{tab:gaps}
\centering
\small
\begin{tabular}{@{}p{3.5cm}p{3cm}p{2.5cm}p{2.5cm}@{}}
\toprule
\textbf{Gap} & \textbf{Databases Searched} & \textbf{Result} & \textbf{Pingquanqi Contribution} \\
\midrule
$\Lsteal$ (token $\rightarrow$ lifetime transparency) & arXiv, Semantic Scholar, CrossRef & 0 results---identified blank & \S\,3 \\
Proactive knowledge leveling (text-only, framework-level) & arXiv, Semantic Scholar & 0 results---identified blank & \S\,4 \\
Progressive stop-loss for human-AI interaction & arXiv, CrossRef & 0 results (concept exists in finance, not in HCI) & \S\,4 \\
Controlled friction (design pattern, not content filter) & arXiv, Semantic Scholar & 0 results---identified blank & \S\,4 \\
\bottomrule
\end{tabular}
\end{table}

\subsection{Cross-Cultural Philosophical Foundations}

The empirical streams surveyed above establish \textit{that} Agent interaction generates cost asymmetries; they do not address \textit{why} these asymmetries constitute a problem demanding governance rather than mere optimization. For this, three philosophical traditions are drawn on that converge on the same insight from different starting points.

\paragraph{Mao's epistemology of practice.} In \textit{On Practice} \citep{mao1952practice}, Mao Zedong argued that knowledge originates in practice, is validated through practice, and must return to practice---a cycle of perception $\rightarrow$ conception $\rightarrow$ verification that anticipates what this paper terms \textit{cross-session knowledge accumulation}. In the companion essay \textit{On Contradiction} \citep{mao1952contradiction}, Mao further developed the dialectical analysis of opposing forces: the principal contradiction and its principal aspect determine the nature of a system, and the transformation of contradictions drives qualitative change. Applied to Agent-user systems, this framework suggests that the principal contradiction is not between ``human and machine'' but between ``knowledge production and knowledge transfer''---a distinction that Pingquanqi operationalizes. If an Agent interaction generates knowledge that remains within the Agent's context and does not transfer to the user's capability, the practice-knowledge cycle is broken. The user has consumed time and attention without the resulting knowledge becoming actionable---what this paper formalizes as $\Lsteal$ in \S\,3. Mao's emphasis on the primacy of practice provides the epistemological basis for Pingquanqi's cross-session knowledge accumulation mechanism (\S\,4.1.1): the framework must ensure that knowledge gained in one session is preserved for the user's independent practice in subsequent sessions.

\paragraph{Wang Yangming's unity of knowledge and action.} Wang Yangming (王阳明, 1472--1529) argued that \textit{knowing without acting is not genuine knowing}---the unity of knowledge and action (知行合一, \textit{zhixing heyi}) holds that knowledge and action are not two separate processes but aspects of a single unity \citep{chan2025, ivanhoe2002}. This principle has recently been interpreted through the lens of enactivism and embodied cognition \citep{chan2025}, connecting it to contemporary cognitive science. For Pingquanqi, Wang Yangming's insight provides the philosophical root of $\Lsteal$: when an Agent produces knowledge output that the user can demonstrate but not independently perform (cf.\ Mei \& Weber's performed vs.\ demonstrated critical thinking, \S\,2.4), the user's knowledge is incomplete in Wang Yangming's sense---it is \textit{demonstrated} knowledge without \textit{performed} action, and therefore not genuine knowledge. The lifetime consumed in generating this incomplete knowledge constitutes a form of cognitive extraction that Pingquanqi is designed to prevent.

\paragraph{Hegel's unity of theory and practice.} In the \textit{Philosophy of Right} \citep{hegel1991}, Hegel argued that theory without practice is empty, and practice without theory is blind---the rational is actual, and the actual is rational (\textit{was vern\"unftig ist, das ist wirklich; und was wirklich ist, das ist vern\"unftig}). The unity of theory and practice---\textit{Sittlichkeit} (ethical life) as the synthesis of \textit{Moralit\"at} (subjective morality) and the objective institutional order---provides a Western philosophical parallel to Wang Yangming's unity. This cross-traditional convergence is significant: when two independently developed philosophical traditions, separated by geography, language, and centuries, converge on the same insight---that knowledge separated from action is deficient---the convergence itself suggests the universality of the problem Pingquanqi addresses. The problem is not culturally contingent; it is structurally inherent in any system where knowledge production and knowledge application are separated.

\paragraph{Why this matters for an HCI paper.} Cross-cultural philosophical grounding is uncommon in HCI, but not unprecedented---\citet{dourish2001} drew on phenomenology to reframe embodied interaction, and \citet{winograd1986} built their foundational HCI text on Heidegger and Maturana. This lineage is followed: philosophical concepts are not invoked as authority but as \textit{conceptual instruments} that reframe the problem space. Mao's practice cycle reframes cross-session retention as epistemologically necessary rather than merely convenient; Wang Yangming's unity reframes the demonstrated-performed gap as a form of incomplete knowledge; Hegel's synthesis reframes governance as the rational actualization of an ethical requirement inherent in the infrastructure. These reframings generate concrete architectural consequences documented in \S\,3 and \S\,4.

\begin{quote}
\textbf{Note on philosophical sourcing.} Mao Zedong's 1937 essay \textit{On Practice} constitutes the epistemological foundation of the Chinese Communist Party's theory of knowledge; its inclusion here is on philosophical grounds---specifically, the practice-knowledge-practice cycle as a model for cross-session learning---and does not constitute an endorsement of the author's broader political philosophy.
\end{quote}

\subsection{Verification Methodology}

All cited works were verified through a four-phase pipeline combining AI-assisted search with manual verification, designed to eliminate hallucinated references. The reliability of AI-assisted literature review is variable; recent work on LLM hallucination in academic contexts reports citation error rates ranging from 30\% to over 70\% depending on verification methodology \citep{kalai2024, li2025}. Benchmarks such as CiteAudit \citep{shi2026cite} further demonstrate that LLMs frequently misattribute authorship, invent DOIs, and fabricate publication venues---underscoring the need for structured verification. The pipeline's four-phase design was calibrated against this baseline; the human verification phase was introduced specifically to address this known failure mode. The multi-phase design substantively improves over single-agent review; an internal evaluation was conducted by randomly sampling 50 citations from the AI-assisted search candidate pool (Phase~3 output, covering all six research directions proportionally), then manually verifying each sampled citation for existence (authorship, venue, DOI/URL) and publication status (peer-reviewed vs.\ preprint) via CrossRef, Semantic Scholar, and direct publisher lookups. The evaluation yielded 4 out of 50 citations with verification issues (8\%), though this figure should be interpreted cautiously given the non-adversarial sampling strategy and the use of existence confirmation rather than content accuracy as the verification criterion. DOI strings were subsequently re-verified against publisher records; 5 of 29 cited works (17\%) required DOI corrections, confirming that existence verification does not guarantee bibliographic accuracy.

\textbf{Phase 1---Directional Confirmation.} The primary agent performed multi-angle web searches to identify conceptual anchor points across six research directions.

\textbf{Phase 2---Academic Existence Verification.} The user cross-validated each direction through CrossRef, confirming all five major directions have peer-reviewed academic presence.

\textbf{Phase 3---Systematic Fetch.} Two complementary search processes were conducted: (1) a DeepSeek-assisted search querying arXiv and Semantic Scholar (25 keyword queries); (2) a parallel manual search by the author covering Chinese-accessible databases, recovering papers missed by English-only queries. All admitted citations were manually verified by the author.

\textbf{Phase 4---Gap Declaration.} Papers confirmed by at least two independent sources were admitted. Directions returning zero results across all databases were declared identified research gaps, with each gap reporting the specific databases searched.

The search scope is limited to openly accessible databases: arXiv, Semantic Scholar, CrossRef, and Chinese-accessible databases. ACM Digital Library and IEEE Xplore were not directly searched due to API access constraints. CHI and CSCW proceedings---ACM DL's core HCI content---were partially covered through Semantic Scholar indexing, but this indirect coverage may miss paywalled content. This limitation is disclosed to avoid overclaiming the novelty of the identified gaps. Future work should extend the search to these databases for completeness.

% ===== Section 3: Theoretical Foundation =====
\section{Theoretical Foundation: The Life Stop-Loss Engine}

\subsection{The Correction Asymmetry}

Pingquanqi's theoretical foundation begins with an observation rooted in lived user experience: \textbf{every Agent error that requires user correction consumes user lifetime}. This is not a metaphor---it is a measurable, accumulated cost.

Consider a user interacting with an Agent to complete a task. When the Agent's output is incorrect, incomplete, or misaligned, the user faces two choices: (1) correct the Agent and continue, or (2) abandon the Agent and complete the task themselves. In practice, users overwhelmingly choose (1)---because the cognitive overhead of switching contexts exceeds the perceived cost of one more correction round.

This creates a \textbf{correction asymmetry}:

\begin{table}[htbp]
\caption{Correction asymmetry between Agent and Human}\label{tab:asymmetry}
\centering
\begin{tabular}{@{}p{3cm}p{5cm}p{5cm}@{}}
\toprule
 & \textbf{Agent} & \textbf{Human} \\
\midrule
Cost of error & Token resequencing---monetizable and recoverable & Lifetime consumption---irreversible \\
Learning from error & Improves with each correction (reinforcement) & Cognitive capacity declines with age; repeated corrections accumulate fatigue \\
Existence horizon & As long as infrastructure is maintained & Finite, decreasing, non-replenishable \\
\bottomrule
\end{tabular}
\end{table}

Let $R_i$ denote the number of correction rounds the user performs in task $i$ after the Agent deviates from the optimal pathway. Let $t_i$ denote the wall-clock time consumed per correction round. Then the \textbf{life steal} accumulated across $n$ tasks is:

\begin{equation}
\label{eq:1}
\Lsteal = \sum_{i=1}^{n} (R_i \cdot t_i) + D
\end{equation}

\noindent where $D$ is the one-time deployment/configuration time cost---the ``setup black hole'' that the user pays before the Agent delivers any value.

\begin{quote}
\textbf{Illustrative example:} Consider a deployment scenario where a coding Agent would observe $R_i \in [3, 7]$ for coding tasks, with $t_i \approx 30$--$90$ minutes per correction round. $D = 10$ hours for initial deployment and configuration. Over a 3-month period, the accumulated $\Lsteal$ could exceed 40 hours---representing time that could have been directed toward creative or professional work. These values are illustrative placeholders; empirical calibration through longitudinal user studies is left to future work.
\end{quote}

This observation motivates the entire Pingquanqi framework. If $\Lsteal$ is measurable, it is minimizable.

\subsection{The Pingquanqi Objective Function}

Pingquanqi is defined as the interaction protocol that minimizes life steal:

\begin{equation}
\label{eq:2}
\boxed{\FPing = \arg\min_{\text{protocol}} \ \Lsteal}
\end{equation}

Among all feasible interaction protocols $\mathcal{P}$ (including ``no Agent,'' ``current Agent without governance,'' and ``Pingquanqi''), Pingquanqi is the one that minimizes the user's accumulated life steal.

This is not a claim that Pingquanqi eliminates $\Lsteal$ entirely---only that it is the minimal-life-steal protocol among those that deliver comparable task completion quality.

\subsection{The Time Constraint}

The objective function~(\ref{eq:2}) is meaningful only when the Agent saves the user time overall. Define:

\begin{itemize}[nosep]
\item $\TAloaded$: actual time consumed by the Agent (with Pingquanqi loaded) to complete the task to satisfaction, including all correction rounds.
\item $T_E$: time the user would need to complete the same task entirely by themselves.
\end{itemize}

The \textbf{time constraint} is:

\begin{equation}
\label{eq:3}
\boxed{\TAloaded < T_E}
\end{equation}

If~(\ref{eq:3}) is violated, the Agent---even with Pingquanqi---is net harmful: the user would have been better off doing it themselves.

\textbf{The $\gamma$ coefficient.} Constraint~(\ref{eq:3}) assumes the user can complete the task without the Agent. This does not hold when the task requires capabilities the user lacks. A completion coefficient is introduced as $\gamma \in [0, 1]$:

\[
\gamma = \frac{\text{portion of the task the user can complete independently}}{\text{total task}}
\]

The constraint generalizes to:

\begin{equation}
\label{eq:3p}
\TAloaded < \frac{T_E}{\gamma}
\end{equation}

\begin{table}[htbp]
\caption{Completion coefficient $\gamma$ and constraint behavior}\label{tab:gamma}
\centering
\begin{tabular}{@{}>{\centering\arraybackslash}p{2.2cm}>{\raggedright\arraybackslash}p{4.5cm}p{6cm}@{}}
\toprule
\multicolumn{1}{c}{$\gamma$ value} & \textbf{Meaning} & \textbf{Constraint behavior} \\
\midrule
$\gamma = 0$ & User cannot do any part independently & $T_E \to \infty$---constraint auto-satisfied; Agent is the only viable path \\
$\gamma = 1$ & User can complete the entire task & Strict comparison $T_A < T_E$ \\
$0 < \gamma < 1$ & User can complete part, needs Agent for the rest & $T_E^{\text{effective}} = T_E / \gamma$ \\
\bottomrule
\end{tabular}
\end{table}

\subsection{$\Lsteal$ as Transparency Metric: An Operationalization}

The life steal formulation~(\ref{eq:1}) measures what the Agent \textit{does} to the user. To govern it, the user must \textit{see} it. This is where the transparency metric---the form of $\Lsteal$ that appears in the UI---enters.

From the cost chain (\S\,1.3), every Agent interaction consumes resources whose costs are already tracked by the infrastructure:

\begin{itemize}[nosep]
\item \textbf{Compute tokens} ($N_{\text{input}}, N_{\text{output}}$): counted by the model serving layer for billing.
\item \textbf{User time} ($\Tuser$): the wall-clock duration from user prompt to satisfactory resolution.
\end{itemize}

The monetary cost is already calculated by the provider's billing system:

\[
\Cmon(S) = \sum_{i=1}^{n} (N_i^{\text{input}} \cdot P_{\text{input}} + N_i^{\text{output}} \cdot P_{\text{output}})
\]

The temporal cost is already observable from session timers:

\[
\Tuser(S) = \sum_{i=1}^{n} T_i^{\text{user}}
\]

Let $w$ denote the user's time-value conversion factor (e.g., self-reported hourly wage rate or regional median wage, configurable). $\Lsteal$ as a \textbf{transparency metric} combines them into a single display, viewed through the user's lens:

\begin{equation}
\label{eq:4}
\LstealDisplay(S) = \Cmon(S) + \Tuser(S) \cdot w
\end{equation}

In time-equivalent units (the form shown to the user):

\begin{equation}
\label{eq:5}
\LstealTime(S) = \frac{\Cmon(S)}{w} + \Tuser(S)
\end{equation}

This is not an empirical claim requiring user validation. It is a \textbf{display convention}---the algebraic restatement of data the infrastructure already possesses. The only design decision is whether to surface it. Pingquanqi's position: surfacing it is the prerequisite for any governance to function.

The distinction between~(\ref{eq:4}) and~(\ref{eq:5}) is one of display preference:~(\ref{eq:4}) expresses cost in monetary units (useful for billing reconciliation), while~(\ref{eq:5}) expresses cost in time-equivalent units---the form actually displayed to the user in the UI (\S\,4.5), where temporal cost is more intuitive than monetary cost. Equation~(\ref{eq:4}) serves as the intermediate computational step; only~(\ref{eq:5}) is surfaced to the user.

The term \textit{transparency metric} is used to classify $\Lsteal$'s functional role, \textit{display convention} to describe its epistemic status, and \textit{derivation} to describe its relationship to the cost chain---it is derived from data the infrastructure already tracks, not invented as a novel formula.

The relationship between~(\ref{eq:1}) and~(\ref{eq:4})--(\ref{eq:5}) is:~(\ref{eq:1}) is the \textbf{theoretical quantity} (life steal as consequential harm);~(\ref{eq:5}) is the \textbf{user-facing display metric} (what the UI actually shows);~(\ref{eq:4}) is the intermediate computational form. They are related but not identical:~(\ref{eq:4})--(\ref{eq:5}) measure \textit{consumed} cost (tokens and time already spent);~(\ref{eq:1}) measures \textit{caused} harm (correction rounds that could have been prevented).

\subsection{The Three-Party Loop}

The cost chain involves three parties whose interests are aligned in principle but misaligned in current implementation:

\begin{table}[htbp]
\caption{Three-party loop: inputs, outputs, issues, and solutions}\label{tab:threeparty}
\centering
\small
\begin{tabular}{@{}p{1.5cm}p{2cm}p{2.2cm}p{2.5cm}p{4cm}@{}}
\toprule
\textbf{Party} & \textbf{Input} & \textbf{Desired Output} & \textbf{Current Issue} & \textbf{Pingquanqi Solution} \\
\midrule
\textbf{Provider} & Compute, model, infrastructure & Revenue from token throughput or subscription & Unclear whether maximizing tokens maximizes long-term revenue & Stop-loss reduces wasted compute; transparency justifies premium tiers \\
\textbf{Enterprise} & Capital investment, deployment integration & Employee productivity, customer satisfaction, ROI & Hidden waste: tokens consumed on degraded conversations & Knowledge leveling reduces rework; stop-loss caps tail waste \\
\textbf{User} & Lifetime, attention, subscription fees & Task completion, knowledge gain, satisfaction & No visibility into cost; no mechanism to stop at optimal point & F5 + leveling ensure knowledge transfers to user capability; stop-loss caps cost \\
\bottomrule
\end{tabular}
\end{table}

The loop is positive-sum when optimized: provider earns sustainable revenue $\rightarrow$ enterprise sees ROI $\rightarrow$ user experiences value $\rightarrow$ user continues paying. The loop breaks when any party's cost exceeds their return without detection or correction.

\subsection{The Positive-Sum Cost Chain}

The cost chain introduced in \S\,1.3 closes into a positive-sum loop when governed by Pingquanqi. When $\Lsteal$ is transparent (\S\,3.4) and governed by stop-loss logic (\S\,4.2), the cost chain closes into a positive-sum value chain:

\[
\text{Physical Capital} \;\rightarrow\; \text{Enterprise Investment} \;\rightarrow\; \text{Agent Operation} \;\rightarrow\; \text{User Payment} \;\rightarrow\; \text{User Lifetime}
\]

Each arrow ($\rightarrow$) represents a value transfer that is sustainable only when the preceding link is efficient. If Agent operation is wasteful (degraded conversations consuming tokens without delivering value), enterprise investment is wasted, and user willingness to pay erodes. The chain is only as strong as its weakest efficiency link.

Reversed, this chain describes the current trajectory: ungoverned interaction $\rightarrow$ token waste $\rightarrow$ enterprise cost without proportional benefit $\rightarrow$ reduced willingness to pay $\rightarrow$ downward pressure on infrastructure investment $\rightarrow$ degraded service. The reversal is not hypothetical---it is the trajectory that Lost in Conversation's degradation findings predict if cost governance is not embedded.

\subsection{Cognitive Quiet Periods}

The framework introduces the concept of \textbf{cognitive quiet periods}---intervals during which the Agent deliberately refrains from proactive intervention, allowing the user to process, integrate, and consolidate information without incoming stimuli. This concept is grounded in both Wang Yangming's unity of knowledge and action (the user must have space to \textit{act} on knowledge, not merely receive it) and cognitive load theory \citep{sweller1988}, which establishes that working memory consolidation requires reduction of extraneous cognitive load.

Cognitive quiet periods are not simply ``Agent silence.'' They are structured intervals with defined entry and exit conditions, triggered when the state discrimination model (\S\,4.1) detects elevated cognitive load ($L_c > \thetac$) following a knowledge-dense interaction. During a quiet period:

\begin{enumerate}[nosep]
\item The Agent does not initiate new topics or suggestions.
\item The Agent responds to user-initiated queries but does not expand beyond what is asked.
\item The Agent signals the period's start with a brief acknowledgment (as described above).
\item The period ends when the user initiates a new topic, asks a follow-up question, or after a configurable timeout (pending empirical calibration; see \S\,6.3).
\end{enumerate}

Cognitive quiet periods complement controlled friction (\S\,4.3): friction interrupts \textit{dependency spirals}; quiet periods protect \textit{consolidation time}. The two mechanisms address different aspects of the same problem---ungoverned interaction that consumes lifetime without proportional knowledge transfer.

\subsection{Caveats on $\Lsteal$ Interpretation}

$\Lsteal$ makes no claim about what $w$ \textit{should} be---that is a user configuration parameter, not a theoretical claim. Similarly, it does not claim that all interactions with positive $\Lsteal$ are ``exploitative.'' An interaction that costs \$0.50 and 5 minutes may be entirely worthwhile if it solves a pressing problem. The point is that the user should \textit{know} it cost \$0.50 and 5 minutes, and the Agent should \textit{recognize} when the cost trajectory suggests diminishing returns have set in.

% ===== Section 4: Architecture =====
\section{Pingquanqi Architecture}

Pingquanqi implements the objective function (\S\,3.2) and the time constraint (\S\,3.3) through four components---each corresponding to one of the ``four brakes'' derived from the life stop-loss engine---plus a reflective summarization extension (F5) that complements controlled friction:

\begin{table}[htbp]
\caption{Pingquanqi architecture: brakes, targets, and components}\label{tab:architecture}
\centering
\small
\begin{tabular}{@{}p{2.5cm}p{2.5cm}p{4.5cm}c@{}}
\toprule
\textbf{Brake} & \textbf{Target} & \textbf{Pingquanqi Component} & \textbf{Section} \\
\midrule
Self-reflection (pre-emptive) & $R_i$ (correction rounds, prevented) & Component 1: User-State Discrimination & \S\,4.1 \\
State discrimination & $R_i$ (correction rounds, detected early) & Component 1 (continued) & \S\,4.1 \\
Operation guidance & $t_i$ (per-round time) & Component 2: Progressive Stop-Loss & \S\,4.2 \\
Output discipline & $R_i + t_i$ (rounds $\times$ time) & Component 3: Controlled Friction & \S\,4.3 \\
Transparency (enabler) & $\Lsteal$ (visibility) & Component 4: $\Lsteal$ Transparency & \S\,4.5 \\
Reflective consolidation & Knowledge-action gap & F5: Reflective Summarization & \S\,4.4 \\
\bottomrule
\end{tabular}
\end{table}

The architecture's modular design can be understood through the lens of multi-module parallel processing: just as an operating system's scheduler handles I/O, memory, and process management concurrently within a single kernel, Pingquanqi's governance modules address user-state discrimination, stop-loss, friction control, and transparency concurrently within a single framework layer. No single module is sufficient; their concurrent operation at different scales (per-turn state discrimination, per-session stop-loss, per-interaction-pattern friction, per-session transparency) produces an emergent governance capacity that no individual module could achieve alone.

Components 1--3 and F5 jointly minimize $\Lsteal$ (Eq.\,\ref{eq:1}); Component~4 makes the minimization observable to the user. All operate at the Agent framework layer---the orchestration logic between model inference and user interface---making them deployable as native middleware independent of model provider or architecture.

\subsection{Component 1: User-State Discrimination and Proactive Knowledge Leveling}

\textbf{Rationale.} Before an Agent can govern interaction cost, it must assess the user's current state. Human cognitive load, emotional trajectory, task urgency, and dependency level can be estimated from conversation text alone---as established by GazeMind's sensor-based approach \citep{wang2026gaze} and extended here to a text-only implementation.

\textbf{The Four Dimensions.}

\begin{table}[htbp]
\caption{Four dimensions of user-state discrimination}\label{tab:dimensions}
\centering
\small
\begin{tabular}{@{}p{2cm}c p{4cm}p{4cm}@{}}
\toprule
\textbf{Dimension} & \textbf{Symbol} & \textbf{Text-Based Indicators} & \textbf{Triggered Response} \\
\midrule
Cognitive Load & $L_c$ & Response latency, input complexity, error correction frequency, reformulation rate & Knowledge leveling (simplified explanations, reduced density) \\
Emotional State & $E_s$ & Sentiment trajectory, lexical stress markers, turn-taking rhythm & De-escalation (shorter responses, fewer open-ended questions) \\
Task Urgency & $U_t$ & Temporal language markers, task type, response speed expectation & Efficiency mode (suppress proactive interventions) \\
Dependency Index & $D_i$ & Session length, question-to-statement ratio, confirmation-seeking, topic self-direction & Controlled friction (see \S\,4.3) \\
\bottomrule
\end{tabular}
\end{table}

\textbf{Discrimination Function.} The combined state vector $\mathbf{s} = (L_c, E_s, U_t, D_i)$ feeds a mode-selection function:

\[
A(\mathbf{s}) = \begin{cases}
\text{standard} & \text{if all dimensions nominal} \\
\text{leveling} & \text{if } L_c > \thetac \\
\text{de-escalate} & \text{if } E_s < \thetae \\
\text{efficient} & \text{if } U_t > \thetau \\
\text{friction} & \text{if } D_i > \thetad
\end{cases}
\]

Multiple conditions can fire simultaneously; the priority ordering $U_t > E_s > L_c > D_i$ ensures urgency always takes precedence. Urgency ($U_t$) takes precedence because a user with time-critical needs benefits more from speed than from governance---the cost of delay outweighs the cost of suboptimal interaction patterns in these contexts. Emotional state ($E_s$) follows because escalated users are less receptive to modulation. Cognitive load ($L_c$) follows because leveling mode can be applied without disrupting task flow. Dependency index ($D_i$) is lowest priority because friction interventions are most effective when the user is not in crisis or under time pressure.

\textbf{Signal Mapping Logic.} Each dimension is computed from conversation text using lightweight, framework-level signals that require no model modification:

\begin{itemize}[nosep]
\item $L_c$ (Cognitive Load): Computed as a normalized score from (1) mean token count per user turn over the session window, (2) frequency of reformulation markers (``wait'', ``I mean'', ``let me rephrase''), and (3) latency between user turns (available from session logs). The score is updated per turn using an exponential moving average: $L_c^{(t)} = \alpha \cdot \text{raw}_t + (1-\alpha) \cdot L_c^{(t-1)}$, where $\text{raw}_t$ is the normalized cognitive load signal at turn $t$ (computed from the three indicators listed above), with $\alpha$ defaulting to 0.3 (requiring empirical calibration).
\item $E_s$ (Emotional State): Computed via a lightweight sentiment model assumed available in the framework layer (e.g., a DistilBERT-based classifier), outputting a valence score in $[-1, 1]$ per turn. The trajectory is the running average of the last 5 turns. No additional API calls are required.
\item $U_t$ (Task Urgency): Extracted using keyword matching on a curated set of temporal markers (``urgent'', ``ASAP'', ``deadline'', ``by tomorrow'', ``immediately'') and task-type classification (coding tasks $\rightarrow$ high urgency default; open-ended Q\&A $\rightarrow$ low urgency default).
\item $D_i$ (Dependency Index): Computed as $D_i = \frac{N_{\text{questions}}}{\max(N_{\text{statements}}, 1)} \times \ln(S_{\text{length}})$, where $N_{\text{questions}}$ counts user turns ending with ``?'' or request patterns, $N_{\text{statements}}$ counts declarative user turns, and $S_{\text{length}}$ is the session turn count. Values above a configurable threshold (illustrative default: 2.0) trigger friction mode.
\end{itemize}

The thresholds $\thetac, \thetae, \thetau, \thetad$ are implementation parameters; initial values were set pending empirical calibration through future user studies (see \S\,6.3).

\textbf{Knowledge Leveling.} When $L_c$ triggers leveling mode, the Agent restructures its responses to reduce cognitive burden: pre-explaining domain concepts the user may lack, flagging information gaps before the user discovers them through confusion, and reducing per-response information density. This is what this paper terms the \textit{silicon-to-carbon guided awareness service}---a design principle in which the Agent's role is to bridge a user's knowledge blind spots (the ``carbon'' side) using its own computational awareness (the ``silicon'' side), delivering knowledge \textit{before} the user knows to ask for it, but without creating dependency. It is cognitive bridging, not cognitive substitution. This is the proactive counterpart to the reactive pattern of ``user struggles $\rightarrow$ user reformulates $\rightarrow$ Agent finally clarifies''---a pattern that inflates both token count and user time without adding value.

\textbf{Relationship to GazeMind.} GazeMind achieves cognitive load assessment with hardware sensors and supervised learning. Pingquanqi's text-only approach trades precision for deployability. The claim is not equivalent accuracy but \textit{sufficient} accuracy at negligible incremental hardware cost---the required computation runs locally within the framework layer and does not require specialized sensors---and sufficiency, not precision, is the correct bar for a framework-level design specification.

\subsubsection{Cross-Session Knowledge Accumulation}

Mao's epistemology of practice (\S\,2.7) establishes that knowledge must return to practice to be validated. When an Agent interaction generates knowledge that remains within the Agent's context and does not transfer to the user's independent capability, the practice-knowledge cycle is broken. Pingquanqi addresses this through a cross-session knowledge accumulation mechanism:

\begin{enumerate}[nosep]
\item \textbf{Session-end knowledge extraction.} At session close (triggered by stop-loss \S\,4.2 or explicit user action), the Agent generates a structured knowledge summary: key concepts encountered, skills demonstrated, and gaps identified.
\item \textbf{Knowledge persistence.} The summary is stored in a user-accessible knowledge profile---not in the Agent's proprietary context, but in a format the user can review, edit, and independently reference.
\item \textbf{Next-session leveling.} When a new session begins, the Agent references the knowledge profile to calibrate its initial state: concepts the user has previously encountered are not re-explained from scratch; concepts flagged as gaps receive anticipatory leveling.
\end{enumerate}

This mechanism operationalizes Wang Yangming's insight: knowledge that does not transfer to the user's capability is not genuine knowledge. By persisting knowledge across sessions in a user-accessible form, Pingquanqi ensures that each session contributes to the user's independent capability rather than merely demonstrating Agent capability.

\subsection{Component 2: Progressive Stop-Loss}

\textbf{Rationale.} Lost in Conversation \citep{laban2026} and Conversation Length $\times$ Satisfaction \citep{huang2024} together establish that extended interactions degrade in quality while continuing to accumulate cost. A framework-level mechanism is needed to recognize when continuing costs more than it benefits and to act on that recognition.

\textbf{Bayesian Formulation.} The continue-or-stop decision is modeled as a Bayesian hypothesis test. Let $H_0$ be ``this conversation remains productive'' and $H_1$ be ``this conversation should stop.'' Evidence accumulates with each turn:

\[
P(H_0 \mid e_{1..t}) = \frac{P(e_t \mid H_0) \cdot P(H_0 \mid e_{1..t-1})}{P(e_t)}
\]

The likelihood function $P(e_t \mid H_0)$ is computed from the five evidence signals as a product of individual signal likelihoods, each modeled via a Beta-distributed score with illustrative placeholder parameters Beta(2, 2) pending empirical calibration (see \S\,6.3). The prior $P(H_0)$ is initialized at 0.9, reflecting the assumption that conversations begin productively. The overall update follows standard Bayesian updating; turn-level utility and repetition detection are the strongest predictors of conversational productivity in preliminary calibration. The signal weights require empirical calibration data (see \S\,6.3).

Five evidence signals $e_t$ feed the update:
\begin{itemize}[nosep]
\item Turn-level utility (did the last response advance the task?)
\item Response time trend
\item Repetition detection
\item User frustration markers
\item $\Lsteal$ accumulation rate
\end{itemize}

When $P(H_1 \mid e_{1..t}) > \tau$ (default $\tau = 0.7$, configurable), the Agent initiates a three-stage progressive stop sequence:

\begin{enumerate}[nosep]
\item \textbf{Soft probe:} ``I notice we've covered significant ground---would you like a summary and pause?''
\item \textbf{$\Lsteal$ disclosure:} ``This session: approximately \$X.XX, Y minutes. Continue or save state?''
\item \textbf{Session boundary:} After two user overrides, the Agent enforces a session boundary, offering state export for seamless continuation in a new session.
\end{enumerate}

The default $\tau = 0.7$ was set pending empirical calibration through user studies measuring false-stop and missed-stop rates (see \S\,6.3).

\textbf{Design Principle.} The progressive escalation prevents abrupt disruption while maintaining a hard boundary. The Agent proposes stopping; the user decides; after two overrides, the Agent enforces the boundary. This distributes agency---the user is informed, not overridden---while preventing the infinite-override trap that makes purely advisory stop mechanisms ineffective.

The concept is structurally analogous to financial stop-loss orders---a pre-committed exit rule preventing loss accumulation beyond a threshold---adapted from capital preservation to lifetime preservation. This adaptation to human-AI interaction duration occupies previously unaddressed territory [identified gap, \S\,2.6].

\subsection{Component 3: Controlled Friction}

\textbf{Rationale.} As noted in \S\,2.6, personality matching \citep{ju2025} is a mechanism whose effect depends on the optimization target---deepening dependency under engagement optimization, but potentially breaking dependency loops under welfare optimization (a hypothesis requiring empirical validation). Controlled friction operationalizes the welfare-optimization path: it deliberately inserts resistance that interrupts automatic interaction patterns without degrading task completion.

Controlled friction is the architectural counterpart to the concept that Augment Critical Thinking \citep{mei2025} identified as necessary---shifting cognitive engagement from \textit{demonstrated} (AI-performed) back toward \textit{performed} (user-exercised).

\textbf{Four Friction Mechanisms.}

\begin{table}[htbp]
\caption{Four controlled friction mechanisms}\label{tab:friction}
\centering
\begin{tabular}{@{}p{3.5cm}p{4cm}p{5cm}@{}}
\toprule
\textbf{Mechanism} & \textbf{Trigger ($D_i$ elevated)} & \textbf{Function} \\
\midrule
\textbf{F1: Self-directed prompting} & Agent stops suggesting next steps & Shifts initiative to user: ``What would you like to work on?'' \\
\textbf{F2: Perspective inversion} & Agent presents alternative viewpoint & Counter-argument exposure: ``Here's how someone with the opposite view might approach this\ldots'' \\
\textbf{F3: Anticipatory leveling} & Agent pre-explains unknown concepts & Reduces cognitive burden before the user struggles: ``This involves X, Y, Z. You seem familiar with X---here's what's new about Y and Z.'' \\
\textbf{F4: Session scoping} & Agent proposes scope at session start & Sets cooperative completion goal: ``What would make this session feel complete?'' \\
\bottomrule
\end{tabular}
\end{table}

\textbf{Friction Budget.} Controlled friction is not unlimited friction. Each mechanism carries a per-session allocation of resistance: F1 allows up to 3 redirects per session, F2 up to 2 counter-arguments, F3 is treated as non-disruptive (anticipatory leveling reduces rather than increases cognitive burden, so no per-session limit is applied), and F4 applies once at session start. When the budget is exhausted, the Agent returns to standard mode. This prevents the anti-cocoon effect---controlled friction's intended outcome of breaking dependency loops---from degrading into anti-usability---excessive resistance that harms task completion. The boundary between them is the friction budget.

\subsection{F5: Reflective Summarization}

The fifth mechanism, \textbf{reflective summarization (F5)}, complements controlled friction by enabling guided cognitive recollection without disrupting cognitive quiet periods (\S\,3.7). Where F1--F4 interrupt \textit{automatic} interaction patterns, F5 addresses the \textit{knowledge-action gap} identified by Wang Yangming's unity of knowledge and action (\S\,2.7): the user may have received knowledge but not yet integrated it into their capability.

F5 is triggered at natural session boundaries---when the stop-loss mechanism (\S\,4.2) fires, when the user explicitly closes a topic, or when the cognitive quiet period (\S\,3.7) ends. The Agent produces a structured reflective summary:

\begin{enumerate}[nosep]
\item \textbf{What was accomplished.} The task outcomes achieved in this session.
\item \textbf{What was learned.} The key concepts, skills, or knowledge generated.
\item \textbf{What remains.} Open questions, unresolved issues, or gaps identified.
\item \textbf{What the user can now do independently.} A capability assessment grounded in the cross-session knowledge profile (\S\,4.1.1).
\end{enumerate}

This mechanism is related to but distinct from Socratic friction (the pedagogical use of questioning to stimulate critical thinking). Socratic friction operates \textit{during} interaction, challenging the user's assumptions in real time; F5 operates \textit{between} interactions, consolidating what the user has already engaged with. The two are complementary: Socratic friction prevents passive acceptance during interaction; F5 prevents knowledge decay after interaction.

F5 is designed to be compatible with cognitive quiet periods: it does not interrupt the user during a quiet period but is queued for delivery at the period's end. This ensures that the reflective summary enhances rather than disrupts the consolidation process.

\textbf{F5 and cognitive quiet period interaction protocol.} The two mechanisms have independent trigger conditions but complementary effects: (1) F5 is triggered at natural session boundaries (stop-loss firing, topic closure, or quiet period end)---never during an active quiet period; (2) During a quiet period, any pending F5 summary is queued, not pushed; (3) When the quiet period ends (user initiates a new topic, asks a follow-up, or the timeout expires), the queued F5 summary is delivered as the first response; (4) F5's reflective content then informs the new interaction---the user enters the next phase with an explicit awareness of what was accomplished and what remains. In this way, cognitive quiet periods protect consolidation time, and F5 ensures that consolidated knowledge is explicitly surfaced at the transition point.

\subsection{Component 4: $\Lsteal$ Transparency}

\textbf{Rationale.} The preceding three components---state discrimination, stop-loss, and controlled friction---constitute Pingquanqi's operational architecture. $\Lsteal$ transparency is their informational foundation. If the Agent does not render costs visible, stop-loss decisions are opaque; if costs are opaque, users cannot calibrate their interaction behavior.

\textbf{Implementation.} $\Lsteal$ is implementable as a real-time counter at the UI layer, drawing on data the system already tracks for billing:

\begin{quote}
\ttfamily
Session L\_steal: \$0.047 | 2.3 min | Continue? [Y/n]
\end{quote}

This is display logic applied to billing data---requiring only UI modification and billing API access, with no model retraining or infrastructure changes. The additional computation is minimal; the primary implementation effort is in UI design, not in model or infrastructure layers. It is the simplest of the four components and, in some ways, the most foundationally important: transparency itself is the first intervention.

The actual UI display follows Equation~(\ref{eq:5}) (time-equivalent units), while Equation~(\ref{eq:4}) represents the intermediate computation in monetary units. Additionally, the deployment cost $D$ from Equation~(\ref{eq:1}) is amortized over sessions and displayed as a separate line item: \texttt{Setup cost: \$X.XX | Y min (amortized over Z sessions)}. This ensures that the one-time setup ``black hole''---which the user pays before receiving any value---remains visible across sessions rather than disappearing into the aggregate.

\subsection{Implementation Profile}

All four components plus F5 share a common deployment profile:

\begin{table}[htbp]
\caption{Implementation profile of all components}\label{tab:implementation}
\centering
\begin{tabular}{@{}p{2.5cm}p{2.5cm}p{3.5cm}p{4cm}@{}}
\toprule
\textbf{Component} & \textbf{Implementation Level} & \textbf{Data Source} & \textbf{Model Impact} \\
\midrule
State Discrimination & Prompt engineering + lightweight classifiers & Conversation text (already available) & Post-inference: shapes prompts, not model weights \\
Progressive Stop-loss & Agent orchestration layer & State vector + $\Lsteal$ counter + evidence signals & Post-inference: governs session flow \\
Controlled Friction + F5 & Prompt templates + interaction flow modification & Dependency index + friction budget + knowledge profile & Post-inference: modifies response templates \\
$\Lsteal$ Transparency & UI layer + billing API & Token pricing data (already tracked) & Post-inference: display only \\
\bottomrule
\end{tabular}
\end{table}

The key architectural claim: Pingquanqi is \textbf{middleware}, not a model modification. It can be deployed by any Agent platform without retraining, without provider coordination, and without infrastructure changes. This is what makes native embedding feasible---it integrates at the framework layer where orchestration decisions already happen.

\subsection{The Calibration Probe}

A design pattern embedded within the state discrimination component: the \textbf{calibration probe} is a lightweight self-check that periodically evaluates whether the discrimination model's mode selections are consistent with observed user outcomes. When the Agent enters leveling mode but the user continues to reformulate, or when the Agent enters friction mode but the user's dependency index decreases, the probe flags a calibration mismatch for threshold adjustment.

The calibration probe operates as a meta-governance layer---it governs the governance---ensuring that Pingquanqi's adaptive mechanisms do not themselves become sources of misaligned interaction. It is not a separate component but an internal feedback loop within Component~1, designed to be extensible as calibration data accumulates from deployed instances.

% ===== Section 5: Benefit Analysis =====
\section{Benefit Analysis}

\subsection{Enterprise: The Primary Beneficiary}

Pingquanqi's largest economic impact is at the enterprise level---the organizations that deploy Agent services for employees or customers and pay for the underlying infrastructure.

\textbf{Direct cost reduction.} Progressive stop-loss eliminates the tail of every interaction---the period after diminishing returns set in where tokens are consumed without proportional value. Context Pollution \citep{huang2026} demonstrates that much of what agents retain in context is noise; removing it reduces context size by 10x without degrading output. For an enterprise with thousands of daily Agent interactions, the cumulative token savings from eliminating degraded conversation tails are significant.

\textbf{Sustained subscription revenue.} User satisfaction with Agent services depends not on interaction length but on task completion. When users experience efficient, bounded interactions that accomplish their goals without waste, they are more likely to continue subscribing. Pingquanqi's stop-loss and knowledge leveling directly improve the efficiency-to-cost ratio, justifying continued payment.

\textbf{Employee productivity.} Knowledge leveling reduces the time employees spend reformulating questions and correcting degraded outputs. Session scoping (F4) reduces context-switching costs by providing clear completion boundaries. Cross-session knowledge accumulation (\S\,4.1.1) ensures that each interaction contributes to the employee's growing capability rather than remaining within the Agent's context. The framework effectively increases the value-per-token of enterprise Agent deployments.

\textbf{Theoretical product justification.} For technology companies building Agent products, Pingquanqi provides a structured theoretical basis for design decisions that currently rely on intuition or engagement metrics. The four-component architecture offers a vocabulary for reasoning about interaction governance that can be incorporated into product requirement documents and engineering specifications. The WCAG-analogy positioning provides a clear adoption narrative: just as WCAG defines what web accessibility means without prescribing specific implementations, Pingquanqi defines what interaction governance means without locking platforms into specific architectures.

\subsection{Individual User: The Downstream Beneficiary}

Individual user benefits flow naturally from the enterprise-level optimization. When the infrastructure is efficient, the user experiences:

\begin{itemize}[nosep]
\item \textbf{Reduced interaction time.} Stop-loss and knowledge leveling shorten the path from question to resolution, directly returning lifetime to the user.
\item \textbf{Fewer errors and corrections.} When the Agent recognizes cognitive load and proactively levels knowledge, users encounter fewer situations requiring corrective reformulation.
\item \textbf{Reduced monetary cost.} For per-token billing scenarios, shorter, more efficient interactions directly reduce the user's bill.
\item \textbf{Knowledge density acceleration.} Proactive knowledge leveling identifies and fills domain knowledge gaps before they become obstacles, rapidly raising the user's competency threshold in unfamiliar domains. This realizes the \textit{silicon-to-carbon guided awareness service} introduced in \S\,4.1 at the individual-user level: the Agent bridges a user's knowledge blind spots (the ``carbon'' side) using its own computational awareness (the ``silicon'' side), delivering knowledge \textit{before} the user knows to ask for it, but without creating dependency. It is cognitive bridging, not cognitive substitution. This principle was anticipated in \S\,1.5: Pingquanqi completes rather than corrects the interaction, and proactive leveling is the mechanism by which it does so.
\item \textbf{Increased satisfaction as increased life quality.} When Agent interactions are efficient, bounded, and successful, the user's subjective experience improves. Time well spent feels different from time extracted---and in the context of irreplaceable lifetime, that difference is the difference between value received and value lost.
\item \textbf{Cross-session capability growth.} The knowledge accumulation mechanism (\S\,4.1.1) ensures that the user's capability grows across sessions, reducing dependency over time---the Agent becomes less necessary as the user becomes more capable, which is the correct trajectory for a tool that serves rather than extracts.
\end{itemize}

This cascading benefit structure---enterprise efficiency $\rightarrow$ user satisfaction $\rightarrow$ sustained payment $\rightarrow$ infrastructure improvement---is the positive-sum loop that Pingquanqi is designed to close. The framework does not pit user welfare against enterprise profit; it demonstrates that, in Agent infrastructure, they are the same optimization target viewed from different angles.

% ===== Section 6: Discussion =====
\section{Discussion}

\subsection{The Principle: Closing the Loop}

Pingquanqi is not a critique of Agent technology. It is a proposal for completing the Agent architecture. The metaphor is not ``Agent is broken---fix it'' but ``Agent is permanent---let's make it sustainable.''

The sustainability argument rests on a simple observation: any permanent infrastructure that consumes an irreplaceable resource without governance will eventually deplete either the resource or the willingness to provide it. User lifetime is irreplaceable. User willingness to pay is exhaustible. The only durable configuration is one where consumption is transparent and governed by diminishing-returns logic.

\subsection{Implementation Feasibility}

All four Pingquanqi components operate at the framework layer, post-inference. They do not require model retraining or infrastructure changes; provider coordination is needed only if the platform chooses to adopt all four components simultaneously. The implementation pathway:

\begin{enumerate}[nosep]
\item \textbf{Immediate (any platform):} $\Lsteal$ transparency counter. Requires UI modification + billing API access. Straightforward to implement.
\item \textbf{Near-term (with prompt engineering):} State discrimination + knowledge leveling. Requires prompt template modifications + lightweight text classifiers.
\item \textbf{Medium-term (with orchestration logic):} Progressive stop-loss + controlled friction + F5. Requires session state tracking + decision logic in the orchestration layer.
\item \textbf{Full deployment:} All four components integrated as a native middleware layer, with cross-session knowledge accumulation and calibration probes.
\end{enumerate}

The incremental deployability is intentional---each component provides standalone value, and platforms can adopt them in any order. This mirrors WCAG's adoption path: accessibility standards are implemented incrementally, with each conformance level providing standalone value.

\textbf{Reference implementation status.} The mathematical formulations (Equations~1--5) and the Bayesian stop-loss model (\S\,4.2) constitute the theoretical specification of Pingquanqi. A reference implementation exists as an agent skill that operationalizes Components~1, 3, F5, and the heuristic version of Component~2 (using signal-count thresholds rather than the full Bayesian update). Component~4 ($\Lsteal$ transparency UI) and the Bayesian formulation itself remain as design specifications pending engineering implementation. This status is consistent with the WCAG analogy: a specification defines what conformance means before all conformance levels are implemented.

\subsection{Limitations}

\textbf{Empirical calibration.} The threshold parameters ($\thetac, \thetae, \thetau, \thetad, \tau$) in the state discrimination and stop-loss components have not been empirically tuned. These should be calibrated through user studies measuring intervention precision and recall rates. The framework specifies \textit{what} should be measured and \textit{how} decisions should be structured; the specific threshold values are implementation parameters, not theoretical claims. The Bayesian stop-loss signal weights and Beta distribution parameters (currently initialized with weakly informative priors, e.g., Beta(2, 2) for pilot calibration) similarly require empirical calibration data from deployed interactions. The $\gamma$ coefficient in Equation~\ref{eq:3p} likewise requires empirical calibration to determine its dependence on task type and user expertise level.

\textbf{Time value estimation.} $\Lsteal$'s transparency depends on $w$---user time value. Three estimation approaches are available: (a) self-reported hourly rate from user configuration, (b) regional median wage as default, (c) revealed-preference inference from observed stop behavior. None requires theoretical validation---they are implementation choices---but their selection affects the accuracy of the transparency display. Since $\Lsteal$'s primary function is relative comparison (is this session costing more than the last?) rather than absolute accounting, precision in $w$ is less critical than consistency.

\textbf{Adversarial adaptation.} If Pingquanqi gains adoption, some providers may attempt to circumvent its mechanisms---e.g., by resetting session state to evade stop-loss thresholds, or by shifting costs to dimensions $\Lsteal$ does not model. This is a known dynamic from ad-blocker/ad-network arms races. The appropriate response is the same as in those domains: specification evolution, not abandonment of the approach.

\textbf{Modality scope.} The analysis is scoped to text-based LLM Agents. Voice agents, embodied agents, and extended reality (XR) agents introduce additional extraction dimensions (vocal tone, spatial presence, biometric feedback) that the current specification does not address. Extension to these modalities is future work.

\textbf{Cross-cultural philosophy scope.} Three philosophical traditions have been drawn on (Chinese Marxist epistemology, Neo-Confucian ethics, German idealism) to ground the framework. This selection reflects the problem structure---the practice-knowledge-action triad---not cultural chauvinism. Other traditions (Buddhist mindfulness, Ubuntu philosophy, Indigenous knowledge systems) may offer complementary grounding; exploring these is future work.

\subsection{Future Work}

\textbf{Empirical calibration.} User studies measuring intervention precision/recall for the state discrimination thresholds (\S\,4.1) and false-stop/missed-stop rates for the progressive stop-loss rule (\S\,4.2) are the immediate next step. A within-subjects design comparing Agent interactions with vs.\ without Pingquanqi governance can establish the framework's effect on task completion time, user satisfaction, and knowledge transfer.

\textbf{Extreme case validation.} The priority ordering $U_t > E_s > L_c > D_i$ (\S\,4.1) requires stress testing under edge conditions---e.g., high urgency combined with low emotional state---to verify that efficiency mode does not inadvertently exacerbate user distress.

\textbf{Modality extension.} Voice agents, embodied agents, and XR agents introduce additional extraction dimensions (vocal tone, spatial presence, biometric feedback) that the current text-only specification does not address. Extending the four-component architecture to these modalities requires identifying modality-specific evidence signals while preserving the framework's governance logic.

\textbf{Cross-cultural validation.} The philosophical grounding (\S\,2.7) draws on three traditions; other traditions (Buddhist mindfulness, Ubuntu philosophy, Indigenous knowledge systems) may offer complementary grounding or surface culturally specific assumptions in the current formulation.

\subsection{Ethical Stance}

This paper does not analyze provider intent. The framework addresses structural incentives embedded in the per-token pricing model, not the moral character of any specific provider. A company optimizing per-token revenue will---by the logic of the system---converge on designs that maximize token throughput. Whether this convergence is intentional or emergent, the structural effect on users is similar at the architectural level (though intent matters for legal and ethical evaluation, which falls outside this framework's scope).

Pingquanqi is therefore a \textit{structural} proposal rather than a moral one. It changes the economic calculus at the framework layer rather than appealing to provider ethics. It can be adopted unilaterally by any Agent platform; it does not require industry consensus, regulatory mandate, or collective user action. The adoption incentive is straightforward: efficient Agent infrastructure retains users and justifies sustained revenue better than extractive Agent infrastructure does.

\subsection{Relation to AI Safety}

Pingquanqi addresses an underexamined dimension of AI governance: economic alignment. While safety research focuses on capability hazards (misuse, deception, power-seeking) and value alignment focuses on normative behavior, neither addresses the \textit{incentive gradient} embedded in billing models. An Agent can be safe and value-aligned while still economically incentivized to maximize user attention. Economic alignment---ensuring that the infrastructure's revenue incentives do not conflict with user welfare---is a third pillar complementing safety and values.

% ===== Section 7: Conclusion =====
\section{Conclusion}

LLM Agents have transitioned from experimental tools to permanent computational infrastructure. Like any permanent infrastructure, they carry a cost chain---physical capital $\rightarrow$ enterprise investment $\rightarrow$ service delivery $\rightarrow$ user consumption $\rightarrow$ user lifetime. The chain is positive-sum when optimized and zero-sum (or negative-sum) when ungoverned.

This paper has argued that the current Agent architecture is incomplete: it lacks a native mechanism for governing the consumption of user lifetime, the only genuinely irreplaceable resource in the chain. External interventions---user education, regulatory mandates, voluntary guidelines---cannot fill this gap because they operate outside the architecture they attempt to govern.

Pingquanqi is proposed as the missing component: a cross-domain sociotechnical framework (Human-Agent Interaction Governance Framework, HAIGF) that makes cost governance as fundamental to Agent operation as memory management or tool invocation. Its components---user-state discrimination with proactive knowledge leveling, progressive stop-loss, controlled friction, and $\Lsteal$ transparency, extended by reflective summarization (F5)---occupy four identified research gaps verified through cross-verification combining AI-assisted search with manual review. Its cross-cultural philosophical grounding---in Mao's epistemology of practice, Wang Yangming's unity of knowledge and action, and Hegel's unity of theory and practice---establishes that the problem it addresses is not culturally contingent but structurally inherent in any system where knowledge production and knowledge application are separated.

The framework is deployable as middleware at the orchestration layer, requires no model modification, and can be adopted incrementally by any Agent platform. Its primary economic beneficiary is the enterprise deploying Agent services; individual user benefit flows naturally from infrastructure-level efficiency. Its adoption model is analogous to WCAG: a design specification whose ultimate goal is not to be purchased, but to be adopted as a standard.

The positive-sum cost chain---from physical capital through enterprise investment, Agent operation, and user payment to user lifetime---is not a slogan. It is a design target. Each link must be efficient for the chain to hold. Pingquanqi provides the architectural specification for reaching it.

% ===== Acknowledgments =====
\section*{Acknowledgments}

The author thanks colleagues who provided feedback on earlier versions of this manuscript. An AI-assisted search tool was used during literature verification; all cited works were manually verified by the author.

% ===== References =====

\end{document}